\documentclass[twocolumn]{aastex62}
\usepackage[utf8]{inputenc}
\usepackage{amsmath}
\usepackage{graphicx}
\usepackage{subfigure}
\usepackage{lipsum}
\graphicspath{{./}{}}
\submitjournal{AJ}
\accepted{July 28, 2020}

\begin{document}

\title{Discovery of a Nearby Young Brown Dwarf Disk}
\author{Maria C. Schutte}
\email{maria.schutte-1@ou.edu}
\affiliation{Homer L. Dodge Department of Physics and Astronomy, University of Oklahoma, 440 W. Brooks Street, Norman, OK 73019, USA}

\author{Kellen D. Lawson}
\affiliation{Homer L. Dodge Department of Physics and Astronomy, University of Oklahoma, 440 W. Brooks Street, Norman, OK 73019, USA}

\author{John P. Wisniewski}
\affiliation{Homer L. Dodge Department of Physics and Astronomy, University of Oklahoma, 440 W. Brooks Street, Norman, OK 73019, USA}

\author{Marc J. Kuchner}
\affiliation{NASA Goddard Space Flight Center, Exoplanets and Stellar Astrophysics Laboratory, Code 667, Greenbelt, MD 20771}

\author{Steven M. Silverberg}
\affiliation{MIT Kavli Institute for Astrophysics and Space Research, Massachusetts Institute of Technology, 77 Massachusetts Ave, Cambridge, MA 02139, USA}

\author{Jacqueline K. Faherty}
\affil{American Museum of Natural History, 200 Central Park West, New York, NY 10024, USA}

\author{Daniella C. Bardalez Gagliuffi}
\affil{American Museum of Natural History, 200 Central Park West, New York, NY 10024, USA}

\author{Rocio Kiman}
\affil{American Museum of Natural History, 200 Central Park West, New York, NY 10024, USA}
\affil{Department of Physics, Graduate Center, City University of New York, 365 5th Ave, New York, NY 10016, USA}

\author{Jonathan Gagn\'{e}}
\affil{Institut de Recherche sur les Exoplan\`{e}tes, Universit\'{e} de Montr\'{e}al, Pavillon Roger-Gaudry, PO Box 6128 Centre-Ville STN, Montreal QC  H3C 3J7, Canada}

\author{Aaron Meisner}
\affil{NSF's National Optical-Infrared Astronomy Research Laboratory,  950 N Cherry Ave, Tucson, AZ 85719, USA}

\author{Adam C. Schneider}
\affil{School of Earth and Space Exploration, Arizona State University, 781 Terrace Mall, Tempe, AZ 85287, USA}

\author{Alissa S. Bans}
\affiliation{Department of Physics, Emory University, 201 Dowman Drive, Atlanta, GA 30322, USA}

\author{John H. Debes}
\affiliation{Space Telescope Science Institute, 3700 San Martin Dr., Baltimore, MD 21218, USA}

\author{Natalie Kovacevic}
\affiliation{Physics Department, Western Illinois University, 1 University Circle, Macomb, IL 61455, USA}

\author{Milton K.D. Bosch}
\affiliation{Disk Detective Citizen Scientist}

\author{Hugo A. Durantini Luca}
\affiliation{Disk Detective Citizen Scientist}
\affiliation{IATE-OAC, Universidad Nacional de C\'ordoba-CONICET. Laprida 854, X5000 BGR, C\'ordoba, Argentina}

\author{Jonathan Holden}
\affiliation{Disk Detective Citizen Scientist}

\author{Michiharu Hyogo}
\affiliation{Disk Detective Citizen Scientist}
\affiliation{Meisei University, 2-1-1 Hodokubo, Hino, Tokyo 191-0042, Japan}

\author{The Disk Detective Collaboration}

\date{\today}
\begin{abstract}
We report the discovery of the youngest brown dwarf with a disk at 102\,pc from the Sun, WISEA~J120037.79$-$784508.3 (W1200$-$7845), via the Disk Detective citizen science project.  We establish that W1200$-$7845 is located in the 3.7$\substack{+4.6 \\ -1.4}$ Myr-old $\varepsilon$~Cha association. Its spectral energy distribution (SED) exhibits clear evidence of an infrared (IR) excess, indicative of the presence of a warm circumstellar disk. Modeling this warm disk, we find the data are best fit using a power-law description with a slope $\alpha = -0.94$, which suggests it is a young, Class II type disk. Using a single blackbody disk fit, we find $T_{\rm eff, disk} = 521 \; K$ and $L_{\rm IR}/L_{\rm *} = 0.14$. The near-infrared spectrum of W1200$-$7845 matches a spectral type of M6.0$\gamma~\pm$ 0.5, which corresponds to a low surface gravity object, and lacks distinctive signatures of strong Pa$\beta$ or Br$\gamma$ accretion. Both our SED fitting and spectral analysis indicate the source is cool ($T_{\rm eff} = $ 2784--2850\,K), with a mass of 42--58\,$M_{\rm Jup}$, well within the brown dwarf regime. The proximity of this young brown dwarf disk makes the system an ideal benchmark for investigating the formation and early evolution of brown dwarfs. 

\end{abstract}

\section{Introduction}
    
    Brown dwarfs (BDs) are substellar objects that range in mass from the deuterium-burning mass limit, approximately 0.013\,$M_{\rm \odot}$, to the H-burning mass limit, approximately 0.075\,$M_{\rm \odot}$ (e.g. \citealt{chabrier2000}).  Because some young BDs host circumstellar disks whose disk masses follows a similar scaling relation as found for young stars (1\% of the total system mass, \citealt{daemgen2016}), it has been suggested that BDs form through a similar mechanism as stars.  However, as summarized by \citet{luhman2012} and references therein, numerous theories for how BDs form have been suggested, including (1) gravitational compression and fragmentation of larger cores that produce a wide range of masses \citep{bonnell2008}; (2) interactions between fragments in larger cores causing low mass objects to be ejected \citep{reipurth2001}; (3) photoionization from OB stars removing the envelope and disk of low-mass protostars \citep{hester1996}; (4) gravitational fragmentation of high-mass protostellar disk creating low mass companions that are ejected \citep{bate2002}; and (5) fragmentation of gas in molecular clouds that produces cores that will collapse over a wide range of masses \citep{padoan2002}. Since the number of known young BDs with disks remains low, distinguishing between these formation mechanisms is challenging \citep{luhman2012}. Studying BDs across a wide range of ages is key to understanding which mechanism is responsible for forming substellar objects.  
    
    Young BDs with disks have been identified in moving groups and associations having a wide range of ages. In the active star formation region Taurus (age \textless 2\,Myr), \citet{monin2010} found a disk fraction for BDs of $41 \pm 12\%$ and a similar disk fraction was found for low mass stars in the region. In similar star forming regions like IC 348 and Chamaeleon I, \citet{luhman2005} found disk fractions consistent with the results from Taurus. \citet{dawson2013} identified 27 BDs with disks out of the 116 objects (23\%) originally selected in the $11 \pm 3$\,Myr \citep{pecaut2012} Upper Scorpius association, a similar disk fraction as found for low mass stars in the same association. These similar disk fractions led \citet{dawson2013} to propose that the disk lifetimes for BDs must also be on the order of 5--10\,Myr. \citet{boucher2016} later found substellar objects with disks in the 10 $\pm$ 3\,Myr \citep{bell2015} TW Hya association \citep{kastner1997,delareza1989}. \citet{boucher2016} also found two substellar objects with disks in the Columba and Tucana-Horalogium associations  \citep{torres2000,torres2008,zuckerman2001,zuckerman2004}, whose older ages (42$\substack{+6 \\ -4}$\,Myr and 45 $\pm$ 4\,Myr respectively; \citealt{bell2015}) suggest BD disk lifetimes might extend much longer than previously expected.

    While we have measured BD disk fractions for very young moving groups and groups aged $\sim$ 10\,Myr, there is a lack of information about BDs with disks in the 3-9\,Myr age range. A search by \citet{luhman2004} for BDs in a different group with a similar age ($\varepsilon$~Chamaeleontis) found none.

    The Chamaeleon cloud complex is a large, nearby star forming region that has been well studied kinematically \citep{marti2013,murphy2013}. It is composed of three dark molecular clouds, Chamaeleon I, II, and III ($d = 165 \pm 50$\,pc; \citealt{marti2013}) and two foreground associations, $\varepsilon$~Chamaeleontis ($d = 102 \pm 4$\,pc; \citealt{murphy2013}) and $\eta$ Chamaeleontis ($d = 95 \pm 1$ pc; \citealt{mamajek1999}). Although the different associations are near each other, it is unclear if $\varepsilon$  and $\eta$~Cha are physically connected to the three dark clouds \citep{marti2013}. The three dark clouds appear to be in various evolutionary stages as well based on the presence of more embedded objects in Chamaeleon II than Chamaeleon I. Chamaeleon III is the youngest dark cloud in the complex and has not begun to form stars \citep{marti2013}. \citet{marti2013} show that Cha I, Cha II, $\varepsilon$~Cha, and $\eta$~Cha have distinct kinematic properties which enables robust membership to be determined.

  The $\varepsilon$~Chamaeleontis ($\varepsilon$~Cha) association is one of the youngest known associations (3.7$\substack{+4.6 \\ -1.4}$\,Myr) in the Solar neighborhood with around 35 known members. The age of the association was estimated by fitting isochrones to all 35 members in the \citet{murphy2013} study and taking the median of those ages, giving a bulk age of 3.7\,Myr. However, there is dispersion expected amongst the members' ages, which is reflected in the uncertainty of the association's age.
 
  \citet{murphy2013} note that fifteen members of this group are disk hosts, indicating a disk fraction of  $\sim$30\%, consistent with the number expected from an association this young. \citet{murphy2013} has shown that the spectral types of the objects in $\varepsilon$~Cha range from K to mid M with most of the objects falling in the M0--M6 range. Only one substellar object, 2MASS~J11550336$-$7919147, has been previously found in the $\varepsilon$~Cha association, and it is a comoving $\sim 10 M_{Jup}$ companion (2M1155$-$79B) to the star 2MASS~J11550485$-$7919108 \citep{dickson2020}. 2M1155$-$79B may instead be a mid-M star occulted by a near edge-on disk (D.A. Dickson-Vandervelde, pvt. comm.).

    \citet{ricci2014} studied three bright BDs with disks in the nearby, very young Taurus star forming region with Atacama Large Millimeter/submillimeter Array (ALMA). They were able to resolve the disks of these objects and determine that the disks are relatively large with the smallest disk extending to $\sim$ 70\,AU \citep{ricci2014}. In comparing the disk radii, dust surface density profiles, and the ratio of disk mass to stellar mass with disks around higher mass stars in the same region, \citet{ricci2014} found that they lie within similar ranges, indicating that perhaps formation mechanisms 1 and 5 as mentioned above are more favorable. In contrast, \citet{testi2016} found that for two bright BD disks in $\rho$ Ophiuchus the outer radii of their disks was truncated around 25\,AU. However, there are still only a handful of measurements of BD disk radii, so more of these measurements are needed to fully understand the formation mechanism. 
    
    In this paper, we report the first discovery of a new, late-M type member of the young $\varepsilon$~Cha moving group that has a disk, making it the closest ($\sim$ 100\,pc) BD younger than $\sim$5\,Myr with a disk. This discovery could help us better understand the formation and early evolution of BDs. 
    In Section \ref{sec:analysis}, we detail our new observations for the object and a composite source plus disk model we fitted. In Section \ref{sec:results}, we present our results from the analysis of the new system. Finally, in Section \ref{sec:discuss}, we discuss our results and draw conclusions about the new brown dwarf system.

\section{Data and Observations}\label{sec:analysis}
\subsection{The Disk Detective Citizen Science Project}

The Disk Detective citizen science project \citep{dd2016}, a joint program of NASA and Zooniverse 
\citep{lintott2008}, identifies circumstellar disk candidates in data from NASA’s \textit{WISE} mission via visual inspection of candidates in the AllWISE catalog \citep{allwise2014}.  The project has published hundreds of new well-vetted disk candidates including the first debris disk discovered around a star with a white dwarf companion \citep{dd2016}, twelve disks around candidate members of comoving pairs \citep{silverberg2018}, a variety of disks in young moving groups \citep{silverberg2018,kuchner2020,silverberg2020}, and the oldest white dwarf with a debris disk \citep{debes2019}. Notably, the Disk Detecitve project has discovered several disk-hosting late-type M stars with extraordinarily long disk lifetimes, called ``Peter Pan'' disks
\footnote{In this case ``Peter Pan'' disks refers specifically to M dwarfs with still accreting disks at ages greater than 20\,Myr rather than the description from A. Sicilia-Aguilar.} \citep[see][]{silverberg2016, silverberg2020}.  Disks in this new class show accretion signatures and other indications of abundant circumstellar gas, yet they are located in young associations (Carina and Columba) with ages in excess of 20\,Myr. A database of Disk Detective objects of interest is available to the public at the Mikulski Archive for Space Telescope (MAST)\href{https://blog.diskdetective.org/2019/09/09/the-disk-detective-database/}.

\subsection{Identification of the W1200$-$7845 Disk System}

The Disk Detective team identified WISEA~J120037.79-784508.3, hereafter called W1200$-$7845, as a viable disk hosting candidate system through its infrared (IR) excess and visual inspection of the images of the system. The discovery was made during the team's vetting of candidate sources with large fractional IR luminosities ($L_{\rm IR}/L_{\rm *}$) that were not previously reported as known disk systems in the refereed literature. 

\begin{figure*}[]
\includegraphics[width=\textwidth]{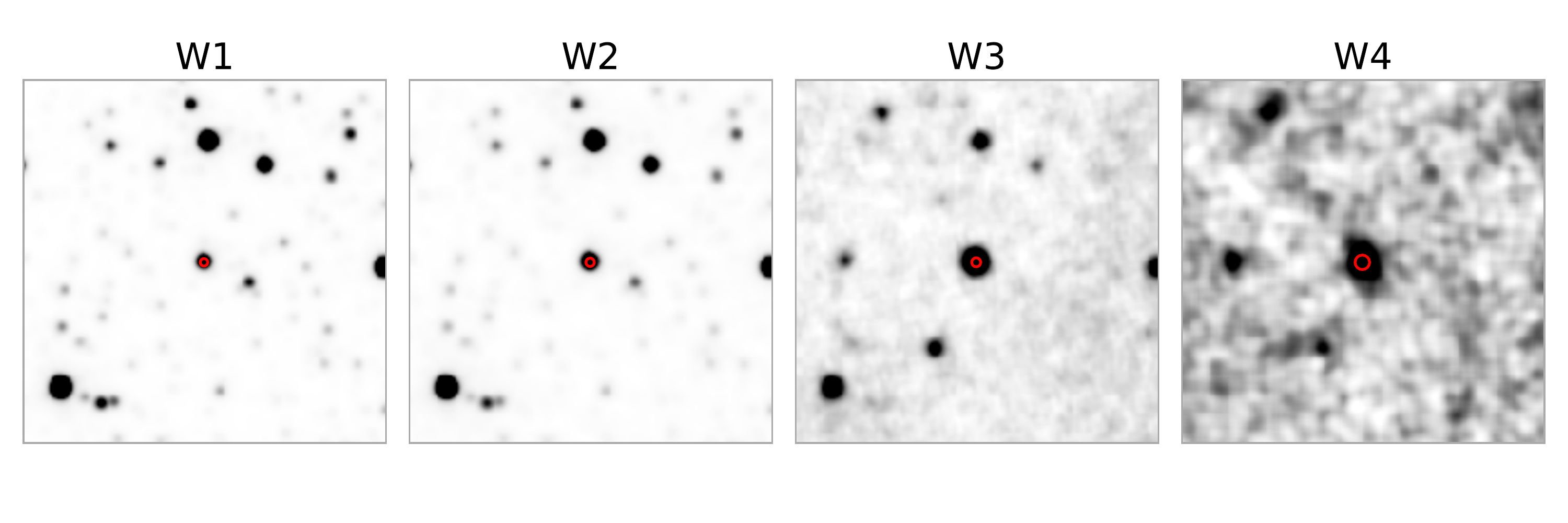}
\caption{WISE images for W1200$-$7845 with the full-width half-maximum (FWHM) of the PSF for each bandpass, 6$\farcs$08, 6$\farcs$84, 7$\farcs$36, 11$\farcs$99 respectively \citep{wise2010}, overplotted in red. The field of view for all images is 300x300 arcseconds. North is up on the image, and east is to the left. We observe a strong IR excess in each WISE bandpass, and we see no particular evidence of source confusion or contamination in any of these bandpasses. 
\label{fig:image}}
\end{figure*}

After initial identification as an object of interest, we further vetted W1200$-$7845.  As seen in the images of the object in Figure \ref{fig:image}, W1200$-$7845 is well separated from nearby sources as there are no additional bright sources within the point spread function (PSF) for each bandpass. 
The other sources in Figure \ref{fig:image} appear to be background sources, and their proper motions \citep{gaia2018} do not place them within any of the PSFs at the time of the WISE observation. From a simple blackbody fit to the spectral energy distribution (SED), its high estimated fractional infrared luminosity,  $L_{\rm IR}/L_{\rm *} = 0.078$, and low estimated stellar effective temperature, $T_{\rm eff} = 2329$\,K, was flagged as being abnormal compared to the bulk Disk Detective disk population. After running W1200$-$7845 using its position, proper motion, and distance \citep{gaia2018} through the BANYAN $\Sigma$ algorithm \citep{banyan2018}, it was identified as a member of $\varepsilon$~Cha moving group with a probability of 99.8\%. 

Since our literature searches revealed W1200$-$7845 had not been previously reported as young disk system, we began more detailed analysis of the system.

\subsection{FIRE Spectrum}

    We obtained a near-infrared spectrum of W1200$-$7845 using the Folded-port Infrared Echellete (FIRE) instrument \citep{fire2008} on the Magellan 6.5-meter telescopes at Las Campanas Observatory. The observations were made in prism mode with the 0$\farcs$6 slit, which achieves a resolving power of $\sim$400 from 0.8 to 2.5\,$\mu$m. Eight 63.4 second exposures were taken in sample-up-the-ramp mode for a total exposure time of $\sim$500 seconds. In the sample-up-the-ramp mode, the detector is constantly read out during the exposure rather than multiple times at the beginning and end of the exposure. This saves on overhead time before and after the exposure, and helps to reduce the read noise for the instrument \citep{fire2008}. The A0-type star HD116270 was observed immediately after W1200$-$7845 for telluric corrections. We reduced the data with a modified version of the SpeXTool package \citep{vacca2003,cushing2004}. Figure \ref{fig:spectrum} shows the reduced spectrum. We note that the conditions during the observation were sub-optimal due to some intermittent clouds.


    \begin{deluxetable*}{@{\extracolsep{3pt}}cccccccccccc}
    \tablewidth{0pt}
    \tablecaption{W1200$-$7845 System Information}
    \tablehead{\multicolumn{4}{c}{} & \multicolumn{8}{c}{Absolute Photometry} \\
    \cline{5-12}
    \colhead{$\mu_{\rm \alpha}\,(\rm mas/yr)$} & \colhead{$\mu_{\rm \delta}\,(\rm mas/yr)$} & \colhead{d (pc)} & \colhead{$\pi$ (mas)}  & \colhead{$M_G^a$} & \colhead{$M_J^b$} & \colhead{$M_H^b$} & \colhead{$M_K^b$} & \colhead{$M_{W1}^c$} & \colhead{$M_{W2}^c$} & \colhead{$M_{W3}^c$} & \colhead{$M_{W4}^c$}}
    \startdata
    -41.543 & -6.120 & 102.32 & 9.77 & 11.325 & 7.468 & 6.946 & 6.545 & 6.159 & 5.564 & 3.682 & 1.99 \\
    $\pm$ 0.153 & $\pm$ 0.126 & $\pm$ 0.96 & $\pm$ 0.09 & $\pm$ 0.020 & $\pm$ 0.033 & $\pm$ 0.030 & $\pm$ 0.031 & $\pm$ 0.031 & $\pm$ 0.029 & $\pm$ 0.031 & $\pm$ 0.08
    \enddata
    \tablecomments{Basic properties of W1200$-$7845, including multi-band absolute photometry, are given.  Note that the photometry is not  extinction corrected, and the full line of sight reddening is E$(B-V)_{\rm S\&F} = 0.1906$ with R$_V = 3.1$.\\
    $^a$ \citet{gaia2018}\\
    $^b$ \citet{2mass2006}\\
    $^c$ \citet{wise2010}}\label{tab:system_info}
    \end{deluxetable*}

\section{Analysis}\label{sec:results} 
\subsection{Confirming Membership}\label{sec:member}
We first identified W1200$-$7845 as a member of $\varepsilon$~Cha through the use of the BANYAN $\Sigma$ algorithm \citep{banyan2018}, which assigned it a 99.8\% membership probability.  Establishing moving group membership generally relies on full three-dimensional kinematic data, and we lack a radial velocity measurement for this source.  However, $\varepsilon$~Cha is known to be well separated from its surroundings \citep{marti2013}, suggesting that this identification could be robust despite the lack of complete kinematic data.

\begin{figure*}[h]
\includegraphics[width=\textwidth]{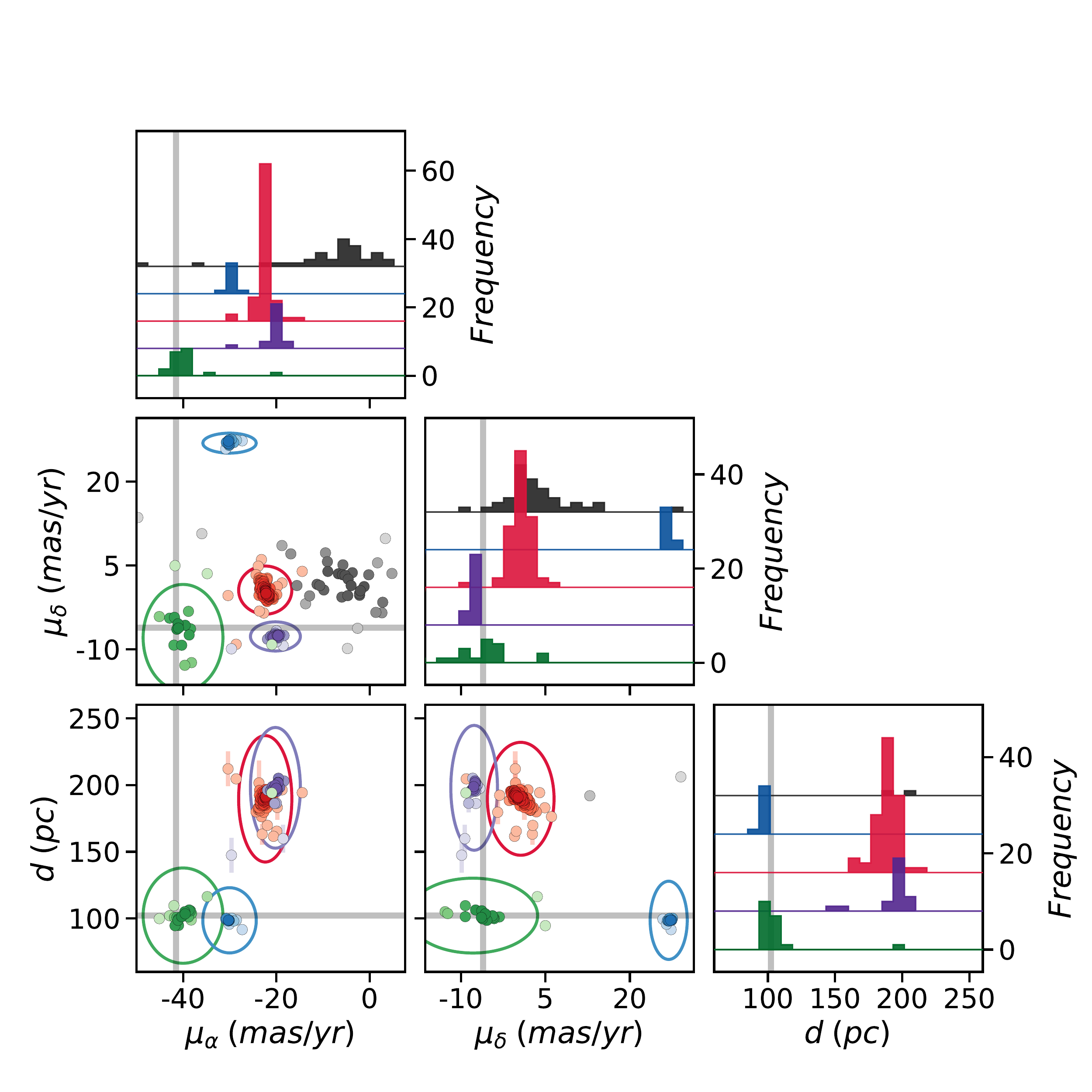}
\caption{
\textit{Gaia} DR2 proper motions and distances for the Chamaeleon Cloud complex \citep{gaia2018}. Colors indicate the association assigned by \citet{marti2013}, including $\varepsilon$~Cha (green), Cha I (red), Cha II (purple), $\eta$ Cha (blue), and background sources (black) (with color saturation in two-dimensional plots corresponding to the local density of that association). Error bars are included for two-dimensional subplots, but are rarely visible. Colored ellipses are drawn qualitatively to highlight the centers of each group. Diagonal elements show histograms for metrics indicated along the x-axis and are offset in the y direction to improve visibility. The majority of background sources fall outside of the distance bounds, which were chosen to better display differences between the various associations. Corresponding values for W1200$-$7845 (gray crosshairs in two-dimensional plots or vertical lines in histograms) demonstrate strong candidacy for membership in the $\varepsilon$~Cha association.
\label{fig:cha_corner}}
\end{figure*}

In order to confirm our preliminary identification of this object as a member of $\varepsilon$~Cha, we re-examined the kinematics of the group.  We started with the same sample of candidate members for $\varepsilon$~Cha, $\eta$ Cha, Cha I and Cha II as \citet{marti2013}, and adopted updated proper motions and distances from \textit{Gaia} DR2 \citep{gaia2018}. We plotted the proper motions and distances against each other for all of the targets, as shown in Figure \ref{fig:cha_corner}.

The results of this analysis show that $\varepsilon$~Cha is well separated from the other three nearby associations in proper motion and distance. W1200$-$7845 has the correct proper motion and distance to put it within 1$\sigma$ of the weighted average of the proper motion for the $\varepsilon$~Cha association, as shown by the crosshairs in Figure \ref{fig:cha_corner}. This result is consistent with the initial BANYAN $\Sigma$ result, which put W1200$-$7845 in the $\varepsilon$~Cha association with a 99.8\% probability. 

\subsection{Dereddening}\label{sec:reddening}
We began our efforts to better characterize the apparent IR excess around W1200$-$7845 by correcting its photometry for attenuation by interstellar dust along the line of sight.  Three-dimensional measurements of interstellar dust reddening (e.g. the Bayestar dustmap of \citealt{green2019}) in the vicinity of W1200$-$7845 are not available as a result of W1200$-$7845's extreme southern declination. While there is a 3D reddening estimation of E$(B-V) = 0.008 \pm 0.017$ in the vicinity of W1200$-$7845 using the STILISM tool \citep{lallement2018}, the grid used is too coarse considering the crowded area of the sky surrounding W1200$-$7845. The STILISM reddening estimation also does not account for the self-extinction that would be significant for a source with a disk like W1200$-$7845. In lieu of this, we adopt the two dimensional reddening at W1200$-$7845's coordinates of E$(B-V)_{\rm S\&F} = 0.1906$ with R$_V = 3.1$ reported by \citet{schlafly2011} as an upper limit to the system reddening. 
    
    \subsection{SED Fitting}\label{sec:modeling}
   Next, we fit W1200$-$7845's SED (Table \ref{tab:system_info}) with models that include the central photosphere flux and the disk flux, which is described either by a power-law, a single blackbody, or two blackbodies. In each case, the composite (source + disk) model also has some fraction of the full line-of-sight extinction applied ($f_{\rm ext}$), with the fraction varying as a free parameter during optimization. To convert A$_V$ to extinction in each bandpass (A$_{\rm f}$), we utilize A$_{\rm f} / {\rm A}_V$ ratios provided by the SVO Filter Profile Service \citep{Rodrigo2012}. 
    
    For the photosphere model, we utilize the solar metallicity BT-Settl-CIFIST grid of photometric evolutionary tracks \citep{baraffe2015}, which are provided in the form of pre-computed absolute magnitudes and are widely used for young, low-mass objects. We did not include the available SkyMapper photometry for W1200$-$7845 since there was no pre-computed absolute magnitudes to compare it to from the BT-Settl-CIFIST evolutionary tracks, and the available DENIS \textit{I} band photometry had twice the error ($\pm 0.045$; \citealt{denis1997}) of the absolute G magnitude from \textit{Gaia}. We also chose to use only the absolute G magnitude from \textit{Gaia} DR2 as the photometric excess factor for W1200$-$7845 for the $G_{\rm BP}$ and $G_{\rm RP}$ filters was 1.689, indicating that W1200$-$7845 has an excess of $\sim 60\%$ from background sources in those filters \citep{evans2018}. Each SED in the model grid is converted to absolute flux for fitting. Our optimization procedure fits to the model grid as provided without any interpolation or re-scaling of the model SEDs. A photosphere model is selected by fitting over a continuous range of source ages and masses, then choosing the value of the nearest discrete point for each parameter from the model grid. Age and mass are used here as the set of models form a clean rectangular grid for these parameters (i.e. the available values of mass are independent of the selected value for age, which is not the case for other parameters). 
   
    The disk component in each case is described by either two parameters (power-law disk and single blackbody disk) or four parameters (two blackbody disk). In both blackbody descriptions, a blackbody disk is described by $T_{\rm eff,disk}$ and $L_{\rm IR}/L_{\rm *}$ (where we optimize for two values of each for the two blackbody disk model\footnote{For this parameterization of the two blackbody disk model the two individual values of $L_{\rm IR}/L_{\rm *}$ give the ratio of each blackbody's luminosity to that of the photosphere. The `true' value of $L_{\rm IR}/L_{\rm *}$ in this description would therefore be attained by summing the  values for the two blackbodies.}). The composite model is formed simply by adding the flux of the photosphere and the blackbody disk(s). For the power-law description, we fit for a power law exponent, $\alpha$, and a logarithmic flux offset, $\log(f_{\rm 0})$.  The solution for this description is piecewise defined, taking the power law values in the WISE filters and the photospheric source values in \textit{Gaia} and 2MASS filters.
  
\begin{figure*}
\includegraphics[width=\textwidth]{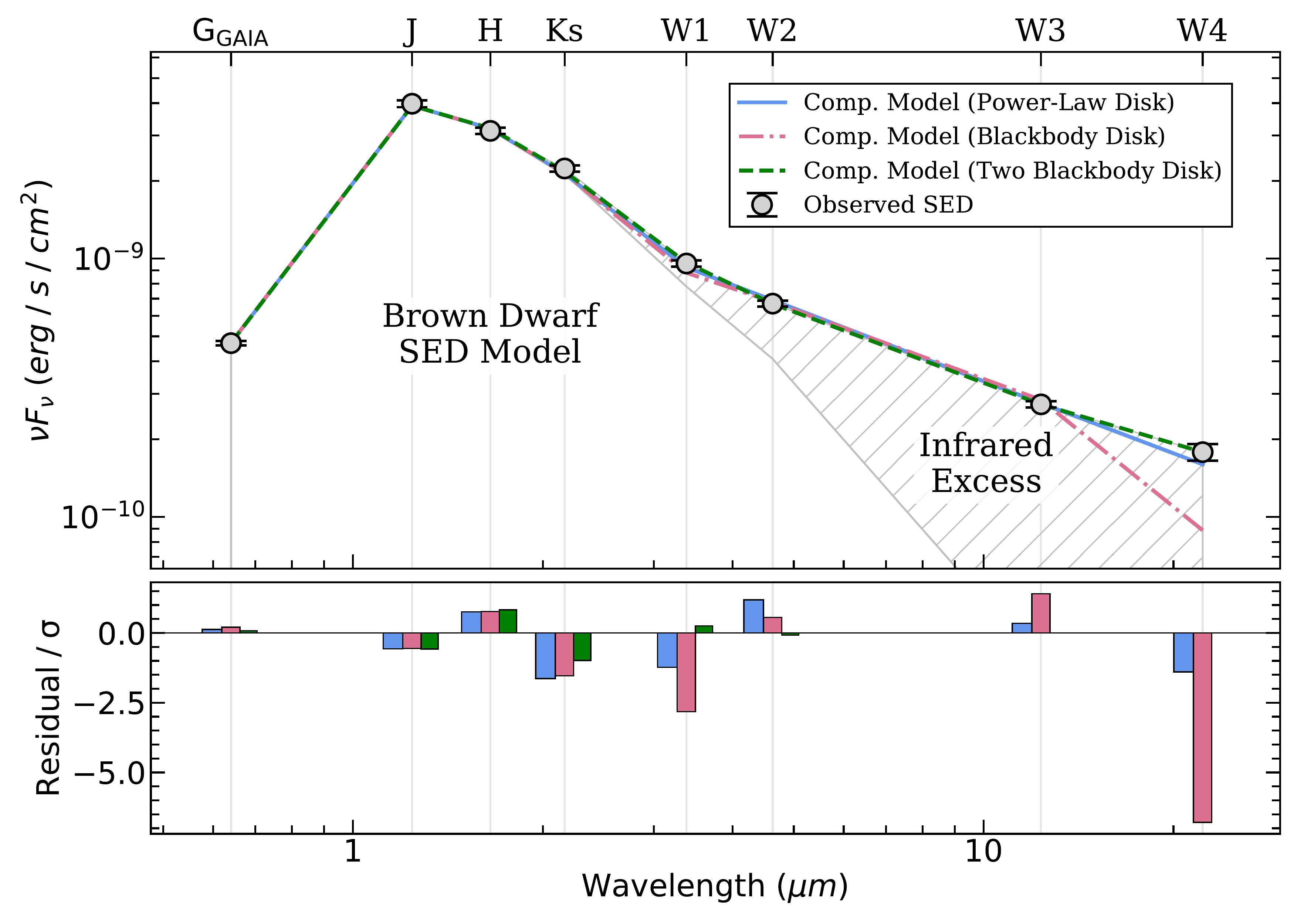}
\caption{The upper subplot shows three different best fit models for W1200$-$7845's SED (grey points) --- a brown dwarf photosphere with a power law disk (blue), a brown dwarf photosphere with a blackbody disk (red), and a brown dwarf photosphere with a disk made of two blackbodies (green). The photosphere profile is taken from utilized evolutionary tracks (Section \ref{sec:modeling}), with no interpolation or rescaling. The fit interstellar reddening has been applied to all models shown. The region of gray hatching indicates the difference between the observed SED and the best-fit photosphere (which is the same for all three types of models). The lower subplot shows the residuals over the photometric uncertainty for the corresponding models in the upper plot. The $\chi^2$ value for the power-law disk model, the single blackbody disk model, and the two blackbody disk model are 9, 60, and 2 respectively.
\label{fig:sedfit}}
\end{figure*}

    Figure \ref{fig:sedfit} shows clear evidence of a warm dust disk component in the \textit{W1-W4} bands shown in the grey hatched region. There is possible evidence of IR excess in the $K_{\rm s}$ band in the grey hatched region that is not connected to the warm dust component, which we discuss further in Section \ref{sec:discuss}. 
    
\begin{figure*}
\includegraphics[width=\textwidth]{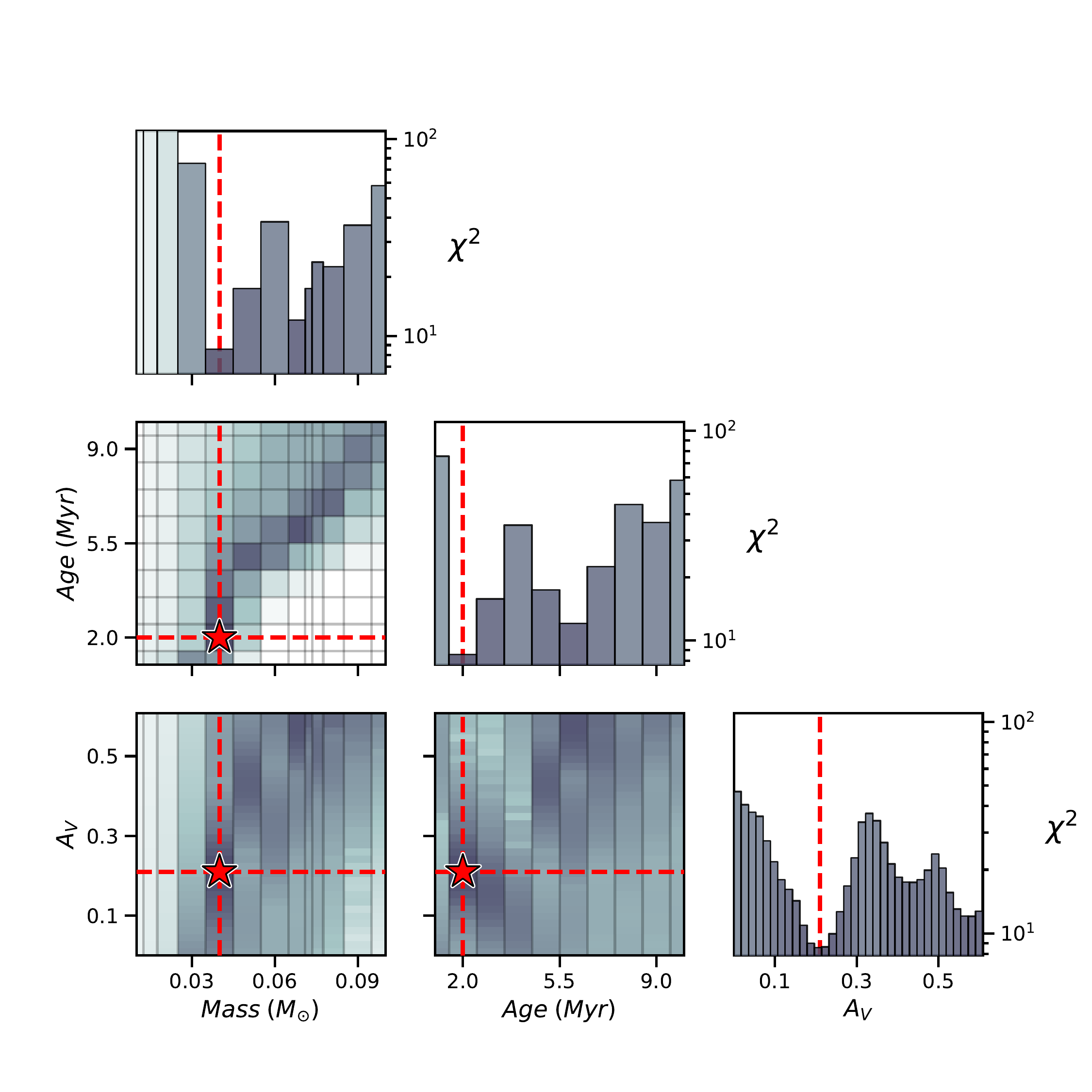}
\caption{A variation of the corner plot for optimization of the composite source + power-law disk model to observations of W1200$-$7845 where the composite model is piece-wise defined, so the disk parameters are independent of the source model parameters. Therefore, the best-fit power-law disk parameters for every possible source model are $\alpha = -0.94$ and $\log(f_{\rm 0}) = -4.79$. Each off-diagonal plot visualizes solutions of two of the three parameters, with each bin colored according to the quality of the best fit achieved with values of the two parameters in that range. Darker bins indicate more optimal solutions. Diagonal elements provide a one-dimensional view of each of the three parameters, indicating the lowest $\chi^2$ value (y-axis) achieved for the binned range of the given parameter (x-axis). For the discrete parameters (source mass and age), bins are drawn according to the resolution available in the model grid; for the extinction A$_V$, the continuous range is divided into 35 equally sized bins. The best-fit solution values are depicted in red (as dashed line cross hairs with a star in 2D subplots, and as a dashed line in 1D subplots).
\label{fig:corner}}
\end{figure*}

    We fit for all parameters (mass, age, $f_{\rm ext}$, and the disk parameters) simultaneously by minimizing the photometric-uncertainty-weighted $\chi^2$. Solutions for each type of composite model are identified using the differential evolution algorithm \citep{Storn1997}, which is well suited for both optimization of discrete parameters and for finding solutions in degenerate parameter spaces. To further investigate possible degeneracies and ensure that the parameter space was well explored, we also iteratively optimized each unique photosphere model of an appropriate age (\textless 10\,Myr) for the remaining parameters (e.g. disk parameters and extinction). The results of fully exploring the parameter space are visualized in Figures \ref{fig:sedfit} and \ref{fig:corner}.

All three types of composite models find the same best-fit parameters for the source component and extinction: $M = 0.04 \; M_{\rm \odot}$ ($42 \; M_{\rm Jup}$), $Age = 2 \; Myr$, and A$_{V} = 0.21$ (see Table \ref{tab:params}). For the single blackbody disk model, we find $T_{\rm eff,disk} = 516 \; K$ and $L_{\rm IR}/L_{\rm *} = 0.14$. For the power law disk model, we find $\alpha = -0.94$ and $\log(f_{\rm 0}) = -4.79$. For the two blackbody disk model, we find $T_{\rm{eff,disk}} = 730 \; K$ and $L_{\rm IR}/L_{\rm *} = 0.09$ for the first blackbody, and $T_{\rm{eff,disk}} = 230 \; K$ and $L_{\rm IR}/L_{\rm *} = 0.06$ for the second. Fit metrics and parameters for each solution are provided in Table \ref{tab:params}. Given that the power law disk model and the single blackbody disk model have the same number of model parameters, the power-law disk appears to be an unambiguously superior explanation of the observed excess, with a $\chi^2$ improvement of $\sim 50$ (see Figure \ref{fig:sedfit}). Assessing the significance of the results for the two blackbody disk model in the context of its increased model complexity is more nuanced. To gauge the merit of additional model parameters, metrics which penalize model complexity, such as the Akaike Information Criterion (AIC) and Bayesian Information Criterion (BIC), are commonly used. However, when the difference between the sample size (n) and the number of free parameters (k) is small (our case), AIC and BIC are not recommended. A more in depth analysis is possible but is beyond the scope of this work. We note, however, that the photospheric parameters identified in the best-fit solution are ubiquitous among the three composite models, lending strong credibility to W1200$-$7845 being a substellar object. The detailed analysis of W1200$-$7845's disk is left to future studies. For the purposes of this paper, we adopt the simpler power-law disk composite model. 

We remark that the model freedom enabled by allowing the extinction to be a free parameter introduces additional risk of degenerate solutions. To attempt to identify other comparable solutions that may exist, we analyze the full array of trial parameter sets from the optimization procedure for the power law disk model. The corner plot of the optimization's parameter space (Figure \ref{fig:corner}) provides some clues regarding the parameter correlations as well as where other plausible solutions may exist. From this, we identify a single additional solution that produces a $\chi^2$ value similar to our best fit; 
all others have $\chi^2$ values at least twice that of the best-fit.

We note that the source mass identified for this alternate solution, $M = 0.07 \; M_{\rm \odot}$, would place the object just below the minimum mass for stable hydrogen burning of $\sim 0.075 \; M_{\rm \odot}$ \citep{chabrier2000}, with the best-fit solution of $M = 0.04 \; M_{\rm \odot}$ falling well below it.
Although the alternate solution has a similar $\chi^2$ value, if we adopt a threshold of $\chi^2_\nu$ for acceptable solutions of $\chi^2_\nu < \chi^2_{\nu, min} + \sqrt{2/\nu}$ (e.g. \citealt{thalmann2014}), no solutions other than the best-fit solution are acceptable. Additionally, we note that all models point to the object being substellar except for one with that model having about 6 times the $\chi^2$ of the best-fit solution.

\subsection{Spectroscopic Properties}

We compared our FIRE spectrum to spectral templates for types M5--L5, including the $\beta$ and $\gamma$ subtypes, which can be indicators of youth \citep{faherty2016}. We again note the observing conditions were sub-optimal during the observation shown in Figure \ref{fig:spectrum}. We show in Figure \ref{fig:spectrum} (a) W1200$-$7845's spectrum versus an M$5.0\gamma$ template, an $M6.0\gamma$ template and an M$7.0\gamma$ template. We determined that the closest fit was to the M$6.0\gamma$ template. As the fit is not perfect, we include an error of $\pm$ 0.5 in the spectral type for W1200$-$7845. In Figure \ref{fig:spectrum} (b), we show W1200$-$7845's spectrum versus a field age M6.0 template, a M$8.0\beta$ template, and the M$6.0\gamma$ template and note the best fit is to the M$6.0\gamma$ object. We further discuss the various spectral subtypes in Section \ref{sec:discuss}. A field age object with spectral type M6.0 $\pm$ 0.5 has an effective temperature of 2850$\substack{+150 \\ -100}$ K, which is consistent with the $T_{\rm eff}$ from the SED fitting \citep{pecaut2013}. While certain spectral types can exhibit significant changes in $T_{\rm eff}$ based on the age of the object, \citet{filippazzo2015} found no significant changes in the $T_{\rm eff}$ values for young and field age M6-type objects. Using the nominal effective temperature and bulk age of the moving group, we can estimate a mass of $\sim 0.055 \; M_{\rm \odot}$ (58\,$M_{\rm Jup}$) using the BT-Settl-CIFIST evolutionary tracks \citep{baraffe2015}. This mass 
is similar to with the mass estimated from the SED-model fitting. As the conditions were sub-optimal during the observation, we could not measure the extinction of W1200$-$7845 from the spectrum with precision, so we instead dereddened our spectrum using the best-fit extinction value from the SED fitting, A$_V$ = 0.21. We also observe no clear signatures of active accretion in the system via emission at either Pa$\beta$ or Br$\gamma$, which we further discuss in Section \ref{sec:discuss}. 

\begin{figure*}
\includegraphics[width=\textwidth]{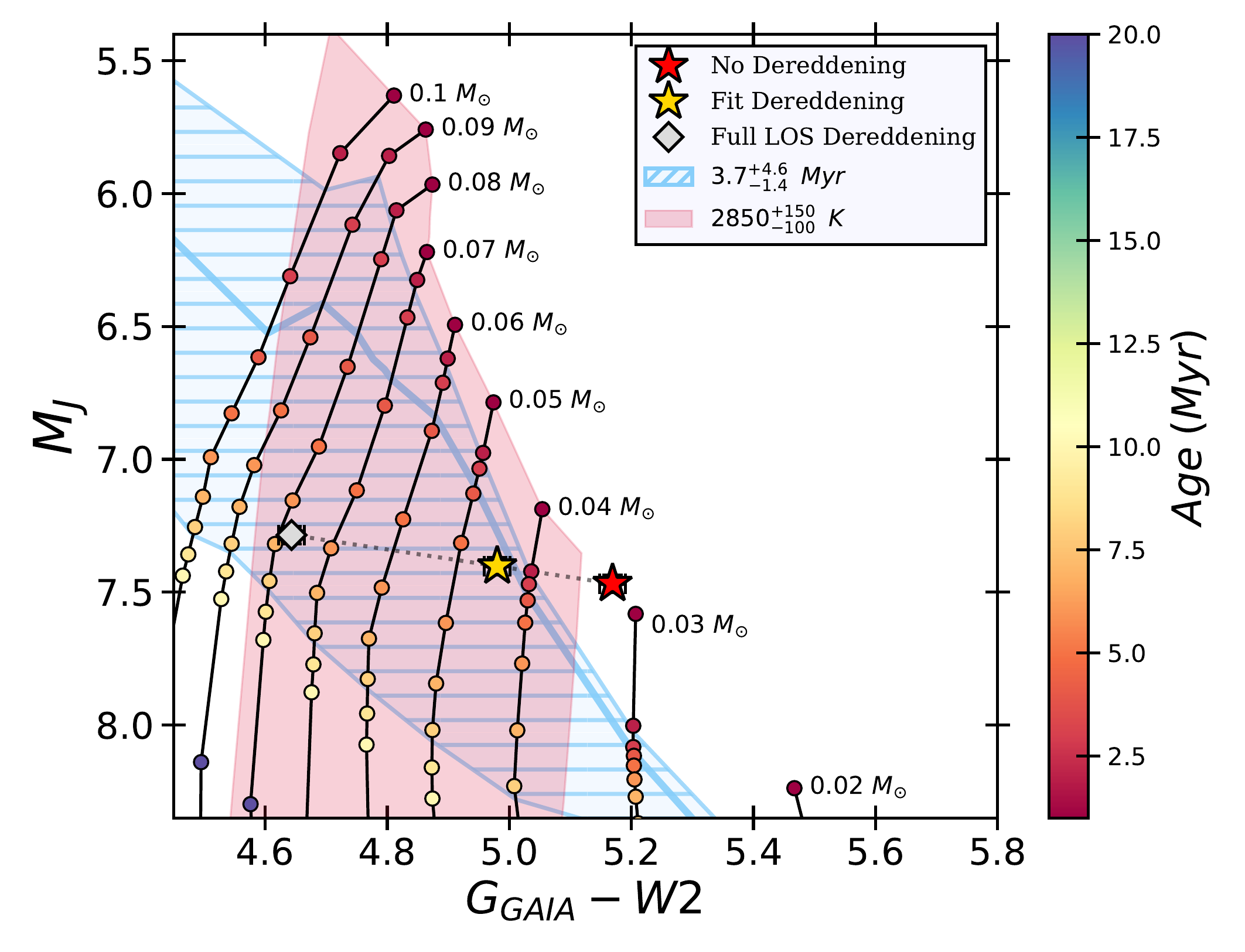}
\caption{$G_{Gaia} - W2$ versus absolute \textit{J} magnitude, showing W1200$-$7845 with no dereddening (red star), fit dereddening (gold star) and full line of sight dereddening (gray diamond), alongside BT-Settl-CIFIST evolutionary tracks (round markers connected by solid lines). The contribution of the disk to W1200$-$7845's SED has been removed by subtracting the fit power law disk profile (see Section \ref{sec:results}). The hatched blue region delineates the $1 \sigma$ age range from \citet{murphy2013} for the association with bounds linearly interpolated from the shown grid of evolutionary tracks. The shaded red region encloses the area of valid $T_{\rm eff}$ within the bounds of the evolutionary tracks and assuming a spectral type of M6.0 $\pm$ 0.5 \citep{pecaut2013}. Considering the range of plausible reddening (see Section \ref{sec:reddening}) and age values, the color-magnitude diagram strongly suggests a substellar mass for the target.
\label{fig:color}}
\end{figure*}

Using the absolute \textit{J} magnitude and the $G_{Gaia} - W2$ color (Table \ref{tab:system_info}), we plotted W1200$-$7845 alongside BT-Settl-CIFIST evolutionary tracks for a mass range of 0.02-0.1  $M_{\rm \odot}$ in Figure \ref{fig:color}. The age range for $\varepsilon$~Cha and the $T_{\rm eff}$ range from the spectral type are plotted as a hatched blue region and shaded red region respectively. If we adopt the dereddening value from our SED fit, W1200$-$7845 falls near the bottom of the $T_{\rm eff}$ range and very close to the median age for the moving group of 3.7\,Myr (shown as thick, solid blue line) with the closest track corresponding to $M = 0.04\,M_{\rm \odot}$.  With no dereddening applied, W1200$-$7845 falls outside both the age and effective temperature ranges, which could be due to a problem with the assumed temperature and age range. If the full line of sight dereddening is applied, W1200$-$7845 is near the top of the $T_{\rm eff}$ range with the closest evolutionary track corresponding to $M = 0.08\,M_{\rm \odot}$. We note that within the given range of plausible reddening, Figure \ref{fig:color} suggests a substellar object.

    
   \begin{deluxetable*}{@{\extracolsep{2pt}}cccccccccccc}
    \tablecaption{W1200$-$7845 SED Fitting Solutions}
    \tablehead{\colhead{} &\colhead{}&\colhead{}& \multicolumn{5}{c}{Photosphere Parameters} & \multicolumn{4}{c}{Disk Parameters} \\
    \cline{4-8}
    \cline{9-12}
    \colhead{Model}&\colhead{$\chi^2$}&\colhead{A$_{V}$}&\colhead{Mass ($M_{\rm \odot}$)}&\colhead{Age ($Myr$)}& \colhead{${T_{\rm eff}} (K)$} & \colhead{Lum. ($L_{\rm \odot}$)} & \colhead{$\log(g)$} & T$_{\rm disk} (K)$ & $L_{\rm IR}/L_{\rm *}$ & $\alpha$ & $\log(f_{\rm 0}$)}
    \startdata
    Blackbody Disk & 59.6 & 0.21 & 0.04 & 2.0 & 2784 & 0.013 & 3.67 & 516 & 0.14 & - & - \\
    Two BB Disk & 2.1 & 0.21 & 0.04 & 2.0 & 2784 & 0.013 & 3.67 & 730, 230 & 0.09, 0.06 & - & - \\
    Power-law Disk & 8.6 & 0.21 & 0.04 & 2.0 & 2784 & 0.013 & 3.67 & - & - & -0.94 & -4.79 \\
    \enddata
    \tablecomments{Best-fit solutions for all three composite models explored (see Section \ref{sec:modeling}). Though only two photosphere parameters are needed for optimization (mass and age are utilized here), we provide additional measures for the best-fit photospheric model for reference.}\label{tab:params}
    \end{deluxetable*}

\section{Discussion}\label{sec:discuss}
We have established that W1200$-$7845 is a likely member of the young ($\sim$3.7\,Myr) $\varepsilon$~Cha association, and that it exhibits an intrinsic IR excess, indicating the presence of a young circumstellar disk. Of the targets from $\varepsilon$~Cha used in our analysis in Section \ref{sec:member} that have radial velocities measured with \textit{Gaia} DR2, the average radial velocity is $14 \pm 5$~${\rm km s^{-1}}$, which is similar to the predicted radial velocity of $14.3 \pm 2.2$~${\rm km s^{-1}}$ from BANYAN $\Sigma$ \citep{banyan2018}. Although radial velocity measurements for W1200$-$7845 do not yet exist, we would expect future measurements of W1200$-$7845's radial velocity to be around $14~{\rm km s^{-1}}$. As $\varepsilon$~Cha is the youngest association within $\sim 100$~pc \citep{banyan2018}, W1200$-$7845 is the youngest BD with a disk currently known within $\sim 100$~pc.

Our best fit SED model gives a power law slope of $\alpha = -0.94$, which would place the disk firmly in the Class II regime by the classification scheme of \citet{lada1987}. Class II objects are akin to the classical T Tauri stars, which are young objects typically with circumstellar disks. As there are not a large number of BDs with disks known, it is hard to determine if Class II BD disks are common, but \citet{downes2015} found that around 20\% of the 15 BD disks in the more distant ($\sim$330\,pc), 7-10\,Myr moving group 25 Orionis were Class II type disks. However, if BDs follow similar formation mechanisms as stars, they should have similar disk properties, disk timescales, and disk fractions as higher mass stars. Our best fit blackbody disk model gives an $L_{\rm IR}/L_{\rm *} = 0.14$, which puts the disk in the primordial disk range given by \citet{cieza2012} as it is greater than 0.1. Debris disks and transitional disks typically have $L_{\rm IR}/L_{\rm *}$ values \textless 0.1 \citep{binks2017}.

\begin{figure}
  \centering
  \begin{tabular}{@{}c@{}}
    \includegraphics[width=.8\linewidth]{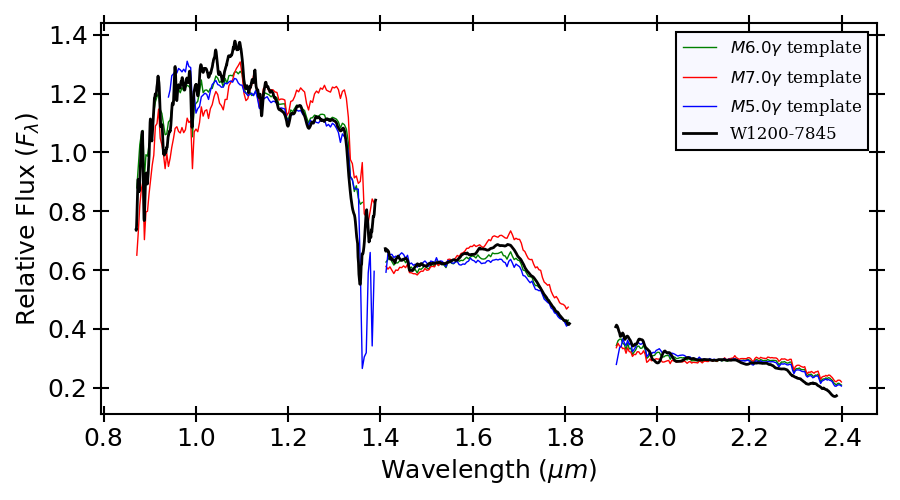} \\[\abovecaptionskip]
    \small (a)
  \end{tabular}

  \vspace{\floatsep}

  \begin{tabular}{@{}c@{}}
    \includegraphics[width=.8\linewidth]{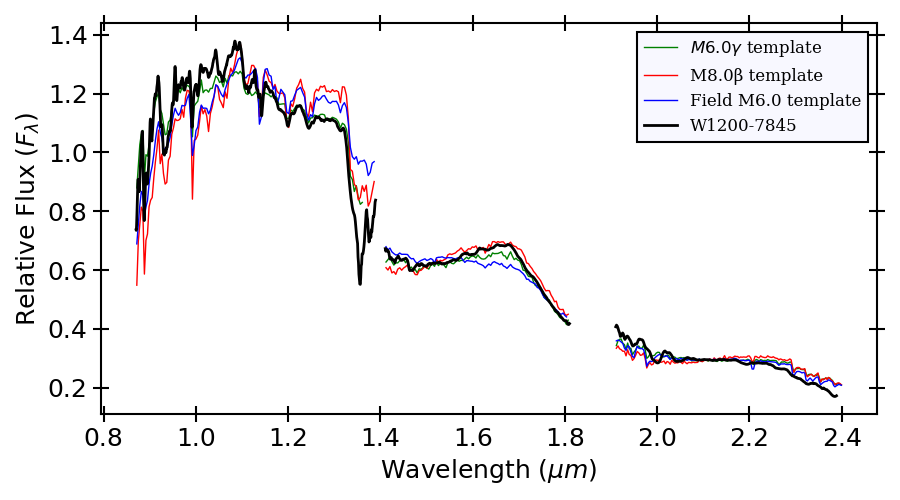} \\[\abovecaptionskip]
    \small (b)
  \end{tabular}

  \caption{(a) Near-IR spectrum of W1200$-$7845 from Magellan FIRE (black) plotted in comparison with M5.0$\gamma$ (blue), M6.0$\gamma$ (green), and M7.0$\gamma$ (red) spectral templates. We adopt a spectral type of M6.0$\gamma$ $\pm$ 0.5 based on visual comparison to the templates, which gives an effective temperature range of 2850$\substack{+150 \\ -100}$\,K \citep{pecaut2013}. There are no signs of strong Pa$\beta$ or Br$\gamma$ accretion present. W1200-7845's spectrum has been dereddened using A$_V$ = 0.21 from the best-fit SED solution.
  (b) Near-IR spectrum of W1200$-$7845 from Magellan FIRE (black) plotted in comparison with M8.0$\beta$ (blue), M6.0$\gamma$ (green), and a field M6.0 (red) spectral templates. W1200-7845's spectrum has again been dereddened using the best-fit extinction of A$_V$ = 0.21 from the SED fitting. The adopted M6.0$\gamma$ type is shown compared to the field M6.0 template and the closest spectral template for the $\beta$ subtype, M8.0$\beta$, as a spectral template for M6.0$\beta$ does not yet exist. The flatness of the H band around 1.65\,$\mu$m in the field age M6.0 spectra is a feature of high surface gravity objects indicating that W1200$-$7845 is likely a low surface gravity object as it does not have that feature. This indicates that W1200$-$7845 is a young object ($< \sim$ 130\,Myr) given the trends found by \citet{faherty2016}. The FIRE spectrum for W1200-7845 shown in panels (a) and (b) is available as the Data behind the Figure.}\label{fig:spectrum}
\end{figure}

Our detailed SED modeling and follow-up spectrum both place W1200$-$7845 firmly in the substellar mass regime.  Our best SED model fit yielded a source $T_{\rm eff}$ of $2784$~K and a mass of $0.04$~$M_{\rm \odot}$ ($42$~$M_{\rm Jup}$). Our follow-up near-IR spectrum of W1200$-$7845 indicated it had a spectral type of M6.0$\gamma$ $\pm$~$0.5$, which has an effective temperature of $2850\substack{+150 \\ -100}$~K \citep{pecaut2013}. 
There are subtypes within the spectral types for brown dwarfs, which can be used as indicators of three gravity classes as described in \citet{faherty2016}. The three classes are field objects, $\beta$ objects, and $\gamma$ objects, which correspond to high surface gravity, intermediate surface gravity, and low surface gravity objects respectively. In comparing W1200$-$7845 to a field age M6.0 object and an M8.0$\beta$ object as seen in Figure \ref{fig:spectrum} (b), W1200$-$7845 does not have the distinct \textit{J}-band features for the two higher surface gravity classes. The $\gamma$ subtype of spectral class we determined for W1200$-$7845 is indicative of a low surface gravity object, which can be an indication of youth (i.e. $< 130$\;Myr) \citep{faherty2016} and supports its membership in a young association. Using the $T_{\rm eff}$ from the spectral type, the age of the group, and BT-Settl-CIFIST evolutionary tracks, we can estimate a mass of $\sim$ 0.055 $M_{\rm \odot}$~(58 $M_{\rm Jup}$).

We examined our near-IR spectrum at both Pa$\beta$ and Br$\gamma$, as these lines have been shown to be accretion indicators for BDs \citep{natta2004}. While other class II objects like classical T Tauri Stars (TTS) do show clear signs of accretion, these rates decrease with source mass \citep{white2001}. This trend extends to the substellar regime, but the typical accretion rates are much lower than for TTS, with values as low as $\sim 10^{-12}\,\mathrm{M_{\odot} yr^{-1}}$ observed \citep{natta2004}. Our non-detection of accretion in W1200$-$7845 could be caused by several factors, the most likely of which is that the accretion rate was too low for us to detect. For very low mass, pre-main sequence stars, \citet{manara2017a} found the minimum amount of accretion needed for detection was $\sim3~{\rm x}~10^{-10}\,\mathrm{M_{\odot} yr^{-1}}$. Moreover, as \citet{nguyen2020} found accretion was sporadic for young very low mass objects; W1200$-$7845 could similarly be accreting sporadically. Additional monitoring of accretion sensitive lines, especially H$\alpha$, are advisable for W1200$-$7845 to help discern between these scenarios. If there is no accretion present, this could imply that the system has begun to clear its inner disk and a cleared or partially cleared inner hole could be present. We note that we do see tentative small excess at the $K_{\rm s}$ band in the SED and such very warm excess must be located close to the host source. If this $K_{\rm s}$ band excess is a real feature, the gas present so close to the source could be source material for future low-level accretion events. Additionally, we remark that we observe no indications that W1200$-$7845's disk is a transitional disk from its infrared SED \citep{espaillat2007t,espaillat2007s}.   

\begin{figure*}
\includegraphics[width=\textwidth]{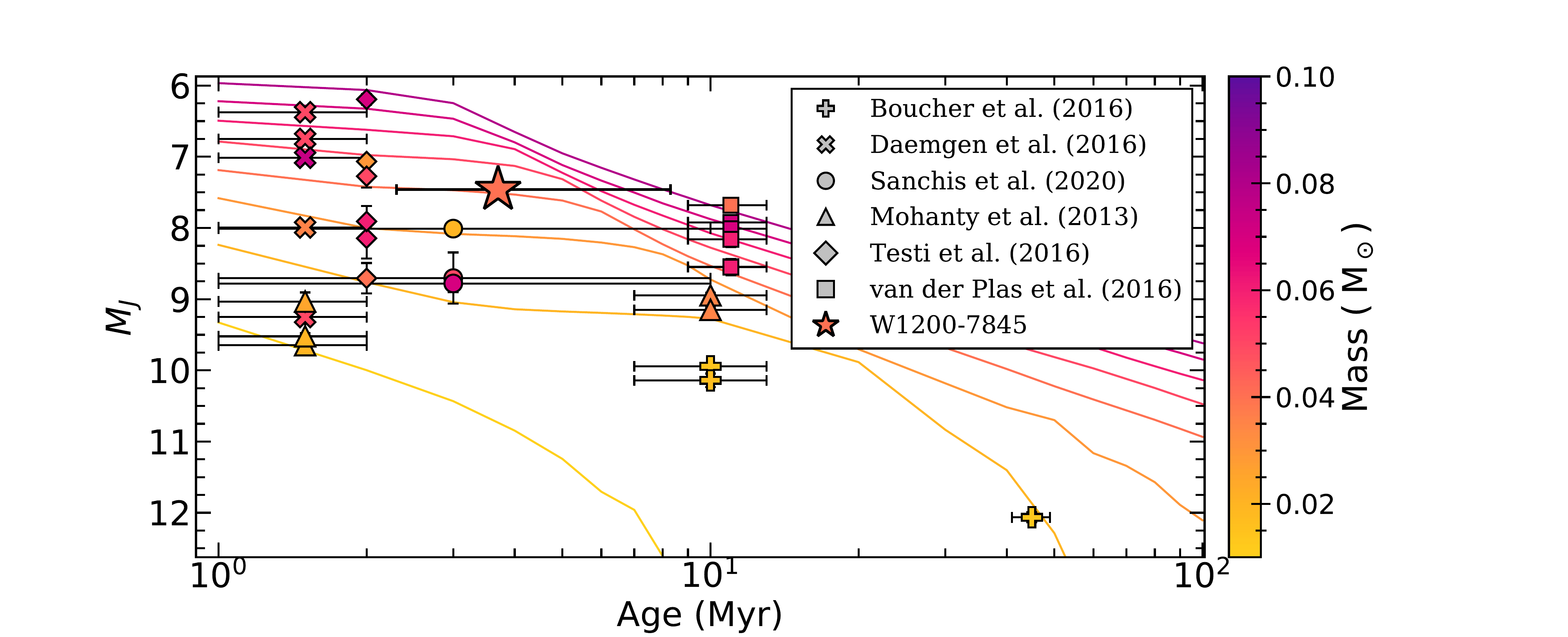}
\caption{ Age versus absolute \textit{J} magnitude for W1200$-$7845 and BDs with disks in the literature that have reported masses, are known members of a nearby ($< 250$\,pc) young moving group, and have a distance from \textit{Gaia} DR2 with lines corresponding to BT-Settl-CIFIST evolutionary tracks \citep{baraffe2015}. The symbols representing brown dwarfs from the literature and the evolutionary tracks are colored according to mass. The selection of objects from \citet{sanchis2020} are young very low mass stars and BDs in the Lupus moving group (3$\substack{+7 \\ -2}$\, Myr; \citealt{alcala2017}). The selection of objects from \citet{daemgen2016} are substellar objects with disks in Taurus (1-2\,Myr; \citealt{kenyon1995}), $\rho$ Ophiuchus ($<$ 2\,Myr), and TW Hya (10 $\pm$ 3\,Myr; \citealt{bell2015}). The objects from \citet{boucher2016} are BDs with disks in TW Hya and the much older Tucana-Horologium group (45 $\pm$ 4\,Myr; \citealt{bell2015}). The objects from \citet{mohanty2013} are more BDs with disks in the Taurus and TW Hya groups. The objects from \citet{testi2016} are BDs with disks from the $\rho$ Ophiuchus moving group, and the objects from \citet{vanderplas2016} are objects from the slightly older Upper Scorpius (11 $\pm$ 2\,Myr; \citealt{pecaut2012}) moving group. 
\label{fig:bd_ages}}
\end{figure*}

We compared W1200$-$7845 to a selection of BDs with disks from the literature in Figure \ref{fig:bd_ages} that spanned an age range of $\sim$ 1--45\,Myr and a mass range of $\sim$ 0.01-0.08\,$M_{\rm \odot}$ ($\sim$ 10-84\,$M_{\rm Jup}$). We selected BDs with reported disks, reported masses, and are in a known nearby (within 250\,pc) young moving group from \citet{sanchis2020,boucher2016,daemgen2016,mohanty2013,testi2016,vanderplas2016} for comparison to W1200$-$7845. While this selection does not include all BDs with disks, it does allow us to compare W1200$-$7845 directly to the other BDs. In plotting absolute \textit{J} magnitude with no extinction correction versus age for these objects with BT-Settl-CIFIST evolutionary tracks in the background, we observe no clear evolutionary trends.  
As is clear in Figure \ref{fig:bd_ages}, W1200$-$7845 provides a nearby probe of BDs with disks in the 3--9\,Myr age range.  While there are BDs with disks in closer associations like TW Hya \citep{boucher2016,schneider2012}, the age of these objects is $\sim 10$~Myr \citep{bell2015}. Increasing the sample size of young BDs with disks as a function of age could help elucidate evolutionary trends of these disk systems.

As W1200$-$7845 is a nearby (102~pc), young ($<5$~Myr) BD with a disk, it has the potential to serve as a benchmark for better understanding the formation and early evolution of BDs. \citet{bate2009} and \citet{bate2012} investigated BDs formed through the ejection scenario \citep{reipurth2001} and found that $\sim 90\%$ of such BDs should have disks truncated at $< 40$~AU. We note that ALMA observations of BDs with disks in the younger, more distant Taurus group have resolved these disks, and found outer disk radii ranging from $\sim$65-140\,AU \citep{ricci2014}. However, \citet{testi2016} were able to resolve two BD disks in $\rho$ Ophiuchus with ALMA and found measured outer radii of $<\,\sim25$ AU, indicating that the disks are truncated rather than extended like the Taurus disks from \citet{ricci2014}. Follow up millimeter observations of the W1200$-$7845 system should be pursued both to quantify radial extent of the disk and to search for a possible cold dust component. 
The former could help determine which formation mechanism is responsible for forming systems like W1200$-$7845.  Moreover, more measurements of the mass and size of BD disks would allow for a more complete understanding of evolutionary trends in such systems, which we previously noted are hard to discern with available data (see Figure \ref{fig:bd_ages}). Follow-up high resolution optical spectroscopy of W1200$-$7845 should also be pursued, as H$\alpha$ is a more robust accretion diagnostic for very low mass objects \citep{natta2004} and thus could be better at identifying any low-level accretion present.

\acknowledgements

We thank our referee for providing us very constructive feedback that helped us to improve the clarity and content of this paper.

We acknowledge support from grant 14-ADAP14-0161 from the NASA Astrophysics Data Analysis Program and grant 16-XRP16\_2-0127 from the NASA Exoplanets Research Program. M.J.K. acknowledges funding from the NASA Astrobiology Program via the Goddard Center for Astrobiology. This publication uses data generated via the Zooniverse.org platform, development of which is funded by generous support, including a Global Impact Award from Google, and by a grant from the Alfred P. Sloan Foundation. This research has made use of the SIMBAD database, operated at CDS, Strasbourg, France.

 Some of the data presented in this paper are available in the Mikulski Archive for Space Telescopes (MAST) at the Space Telescope Science Institute. STScI is operated by the Association of Universities for Research in Astronomy, Inc., under NASA contract NAS5–26555. Support to MAST for these data is provided by the NASA Office of Space Science via grant NAG5–7584 and by other grants and contracts.
 
 This research has made use of the SVO Filter Profile Service (http://svo2.cab.inta-csic.es/theory/fps/) supported from the Spanish MINECO through grant AYA2017-84089.

\bibliographystyle{aasjournal}
\bibliography{apj-jour,references}

\end{document}